\documentclass[twocolumn]{aastex62}

\usepackage{multirow}
\usepackage{amsmath}
\usepackage{chngcntr}

\graphicspath{{./}{figures/}}

\received{}
\revised{}
\accepted{}
\submitjournal{ApJ}

\begin{document}

\title{Monte Carlo Investigation of the Ratios of Short-Lived Radioactive Isotopes in the Interstellar Medium}

\correspondingauthor{Andr\'es Yag\"ue L\'opez}
\email{andres.yague@csfk.hu}

\author[0000-0002-7294-9288]{Andr\'es Yag\"ue L\'opez}
\affil{Konkoly Observatory, Research Centre for Astronomy and Earth Sciences, E\"{o}tv\"{o}s Lor\'and Research Network (ELKH), Konkoly Thege Miklos ut 15-17, H-1121 Budapest, Hungary}
\affiliation{NuGrid Collaboration, \url{http://nugridstars.org}}

\author[0000-0002-9986-8816]{Benoit C\^ot\'e}
\affil{Konkoly Observatory, Research Centre for Astronomy and Earth Sciences, E\"{o}tv\"{o}s Lor\'and Research Network (ELKH), Konkoly Thege Miklos ut 15-17, H-1121 Budapest, Hungary}
\affiliation{ELTE E\"{o}tv\"{o}s Lor\'and University, Institute of Physics, Budapest 1117, P\'azm\'any P\'eter s\'et\'any 1/A, Hungary}
\affiliation{NuGrid Collaboration, \url{http://nugridstars.org}}

\author[0000-0002-6972-3958]{Maria Lugaro}
\affil{Konkoly Observatory, Research Centre for Astronomy and Earth Sciences, E\"{o}tv\"{o}s Lor\'and Research Network (ELKH), Konkoly Thege Miklos ut 15-17, H-1121 Budapest, Hungary}
\affiliation{ELTE E\"{o}tv\"{o}s Lor\'and University, Institute of Physics, Budapest 1117, P\'azm\'any P\'eter s\'et\'any 1/A, Hungary}
\affiliation{Monash Centre for Astrophysics, School of Physics and Astronomy, Monash University, VIC 3800, Australia}

\begin{abstract}

Short-lived radioactive nuclei (SLR) with mean lives below $\sim$ 100 Myr provide us with unique insights into current galactic nucleosynthetic events, as well as events that contributed to the material of our Solar System more that 4.6 Gyr ago. Here we present a statistical analysis of the ratios of these radioactive nuclei at the time of early Solar System (ESS) using both analytical derivations and Monte Carlo methods. We aim to understand the interplay between the production frequency and the mean lives of these isotopes, and its impact on their theoretically predicted ratios in the interstellar medium (ISM). We find that when the ratio of two SRLs, instead of the ratios of each single SLR relative to its stable or long-lived isotope, is considered, not only the uncertainties related to the galactic chemical evolution of the stable isotope are completely eliminated, but also the statistical uncertainties are much lower. We identify four ratios, $^{247}$Cm/$^{129}$I, $^{107}$Pd/$^{182}$Hf, $^{97}$Tc/$^{98}$Tc, and $^{53}$Mn/$^{97}$Tc, that have the potential to provide us with new insights into the \textit{r}-, \textit{s}-, and \textit{p}-process nucleosynthesis at the time of the formation of the Sun, and need to be studied using variable stellar yields. Additionally, the latter two ratios need to be better determined in the ESS to allow us to fully exploit them to investigate the galactic sites of the \textit{p} process.

\end{abstract}



\section{Introduction} \label{sec:intro}

Short-lived radioactive nuclei (SLR) are unstable nuclei with mean lives $\approx$ 0.1 to a 100 Myr. Their abundances can be measured in a variety of locations, both live via $\gamma$-ray spectroscopy \citep{Diehl2010} and analysis of deep-sea sediments \citep{Wallner2015}, and extinct, as in the case of their early Solar System (ESS) abundances inferred through the excess of their daughter nuclei in meteoritic samples \citep{Dauphas2011}. Because of their short mean lives relative to the age of the Galaxy, these nuclei represent the fingerprint of current nucleosynthesis, some of them do not even live enough to travel far away from their site of origin, which results in the decoupling of their abundances from galaxy-wide mixing processes \citep[see, e.g.][]{Diehl2010,Fujimoto2018}.
When considering their evolution in the Galaxy, SLRs therefore probe the current galactic star formation rate instead of the star formation history \citep{Clayton1984,Meyer2000,Huss2009} and, as such, are relatively unaffected by the processes that operate over the full timescale of the Galaxy, such as galactic inflows and outflows (e.g., \citealt{Somerville2015,Naab2017,Tumlinson2017}), the build-up of the total stellar mass (e.g., \citealt{Bland2016}), and the mixing and recycling processes (e.g., \citealt{angles2017}).
Such sources of uncertainty, instead, affect significantly the stable, or long-lived, reference isotope used to measure the abundance of SLR nuclei in the ESS. In \citet{Cote2019} we considered the impact of these sources of uncertainty on the determination of radioactive-to-stable isotopic ratios in the Galaxy and derived that their impact on the ratio results in a variation of at most a factor of 3.5.

There are other sources of uncertainty, however, that must be considered for the evolution of SLRs in the interstellar medium (ISM). As mentioned above, due to their short mean life, SLRs are not evenly distributed in the Galaxy \citep{Fujimoto2018,Pleintinger2019}. In particular, the evolution of a SLR at a specific location in the Galaxy directly depends on the ratio between its mean life $\tau$ and the average time between enriching events $\langle\delta\rangle$, as well as the specific statistical distribution of these $\delta$ \citep[see][henceforth Paper I]{Cote2019B}. The reason for this can be understood by analyzing two limiting cases: $\tau \gg \langle\delta\rangle$ and $\tau \ll \langle\delta\rangle$. In the first case, the mean life is much larger than the time between two enriching events. This allows for the build-up of a memory\footnote{Here we define memory as the SLR abundance remains, non-decayed, from the enrichment events that occurred before the last event.} of the SLR abundance up to a steady-state (between production and decay) equilibrium value equal to the yield of a single event multiplied by a factor $\tau/\langle\delta\rangle$. In the second case, the expected time between two enriching events is instead far apart enough to allow for the complete decay of the SLR before the next event, leaving almost no memory. Therefore, in this case, the average abundance remains below the value of the yield.
In relation to investigations of the ESS, the first case allows us to calculate the isolation time (T$_{\text{iso}}$), defined as the time between the decoupling of the material that ended up in the solar nebula from the Galactic chemical enrichment processes (in other words, the birth of the colder and denser molecular cloud) and the formation of the first solids in the nebula. The second case instead allows us to calculate
the time from the last event (T$_{\text{LE}}$), defined as the time since the last nucleosynthesis event in the Galaxy that contributed a particular SLR to the Solar System matter \citep{Lugaro2014,Lugaro2018}. If T$_{\text{LE}}$ can be calculated, then the SLR may also be used as constraints for the features of specific nucleosynthetic event \citep[see][]{Cote2020}.

In Paper I we analysed the SLR abundance distribution resulting from uneven temporal distribution of nucleosynthetic source, and derived the uncertainties due to this temporal granularity of the enriching events using a simple statistical model of a given region in the Galaxy affected by several enriching events via a Monte Carlo calculation. We concluded that the interplay between the time between two enriching events and the mean life of the SLR determines both the steady-state equilibrium value and its uncertainty. The uncertainty calculated in Paper I does not affect the abundance of the stable reference nucleus, which is well mixed within 100 Myr \citep[e.g.][]{deavillez02}, and can be simply composed with the uncertainty due to the GCE studied by \citet{Cote2019} to calculate the total uncertainty in the SLR/stable isotopic ratio. This total uncertainty can then be used to deduce information about the isolation time (see Paper I, Sect.~5) or the time from the last enriching event \citep[see][]{Cote2020}.

Here, we use the same methodology as in Paper I to study the effect of the presence of heterogeneities due to the temporal granularity of their stellar sources onto the the behaviour and uncertainty of the ratio of two SLRs. Such ratio can exhibit a markedly different behaviour to that of a SLR/stable isotope ratio because its evolution depends also on the difference between the two mean lives. 
We will restrict ourselves to analysing the scenario of $synchronous$ enrichment scenario. That is, the situation in which both SLRs are always generated in the same events. This means that the evolution of the abundances of both isotopes are correlated, and the uncertainty of their ratio cannot be simply derived from adding the individual abundance uncertainties on each isotope. We will also assume that the production ratio $P$ of the two SLRs is always the same. The extension to a more general framework in which different events have different production ratios will not fundamentally change our conclusions, as long as both isotopes are always created together. We do not analyse instead the  complementary $asynchronous$ enrichment scenario, where at least one of the SLR is created in more than one type of event. This scenario is more complex to analyse with our statistical method because it is not possible to define a single production ratio for this case. Furthermore, the possibility that the two SLR may have different $\langle\delta\rangle$ values from different sources complicates the general analysis. 

The outline of the paper is as follows. In Section~\ref{sec:analyticalSolution}, we assume that $\delta$ is constant, and present the analytical solutions to quantify the abundance and uncertainty of any ratio involving two SLRs, for four different regimes. In Section~\ref{sec:stochasticDelta}, we extend our analysis by accounting for a variable $\delta$, and run Monte Carlo calculations to better quantify the uncertainty on SLR abundance ratios. In Section~\ref{sec:discussion}, we apply our statistical framework to radioactive isotopic ratios relevant for the ESS, and discuss the implication of our work on the derivation of T$_\mathrm{iso}$ and T$_\mathrm{LE}$. The codes used in this work are publicly available on GitHub\footnote{\url{https://github.com/AndresYague/Stochastic_RadioNuclides}}.

\section{The case of $\delta$ = $\delta_c$ = constant} \label{sec:analyticalSolution}

We start with the analysis of the simplest case, which assumes that the time between enriching events $\delta$ is constant. The steady-state abundance (in mass) of a single SLR with mean life $\tau$ is
\begin{equation}
    M = M_{\text{ej}}\frac{1}{1 - e^{-\delta_c/\tau}}e^{-\Delta t/\tau},
    \label{eq:evolOne}
\end{equation}
where $M_{\text{ej}}$ is the ejected mass from a single event, $\delta_c$ is the constant time between two successive enrichments, and $\Delta t < \delta_c$ is the time since the last enrichment \citep[see][]{Lugaro2018}.

By taking Equation (\ref{eq:evolOne}) for two isotopes $M_1$ and $M_2$ with mean lives $\tau_1$ and $\tau_2$ respectively, the steady-state evolution of their ratio can be described as:
\begin{equation}
    \frac{M_1}{M_2} = P\,\frac{1 - e^{-\delta_c/\tau_2}}{1 - e^{-\delta_c/\tau_1}}e^{-\Delta t/\tau_\text{eq}},
    \label{eq:evolRatio}
\end{equation}
where $P$ is the production ratio at the stellar source, and $\tau_\text{eq}$ is the \textit{equivalent} mean life given by
\begin{equation}
    \tau_\text{eq} = \frac{\tau_1\tau_2}{\tau_2 - \tau_1},
    \label{eq:tauEq}
\end{equation}
and representing the mean life of the ratio of the radioactive isotopes. Note that $\tau_\text{eq}$ can be negative if $\tau_1 > \tau_2$. Although we consider generally the case where $\tau_\text{eq}$ is positive, we will explain the differences with the negative case, wherever they exist.

The time-averaged value of Equation (\ref{eq:evolRatio}), is given by (see Appendix \ref{sec:mathDevelopment})
\begin{equation}
    \left\langle\frac{M_1}{M_2}\right\rangle = \mu = P\, \frac{\tau_\text{eq}}{\delta_c}\,\frac{1 - e^{-\delta_c/\tau_2}}{1 - e^{-\delta_c/\tau_1}}\,\left(1 - e^{-\delta_c /\tau_\text{eq}}\right),
    \label{eq:avgRatio}
\end{equation}
and the difference between its maximum and minimum (derived by taking $\Delta t = 0$ and $\Delta t = \delta_c$ in Equation (\ref{eq:evolRatio})) values can be written as
\begin{equation}
    \text{Max} - \text{Min} = \mu\frac{\delta_c}{\tau_\text{eq}}.
    \label{eq:maxMinRatio}
\end{equation}

\begin{figure*}
    \includegraphics[width=7.0in]{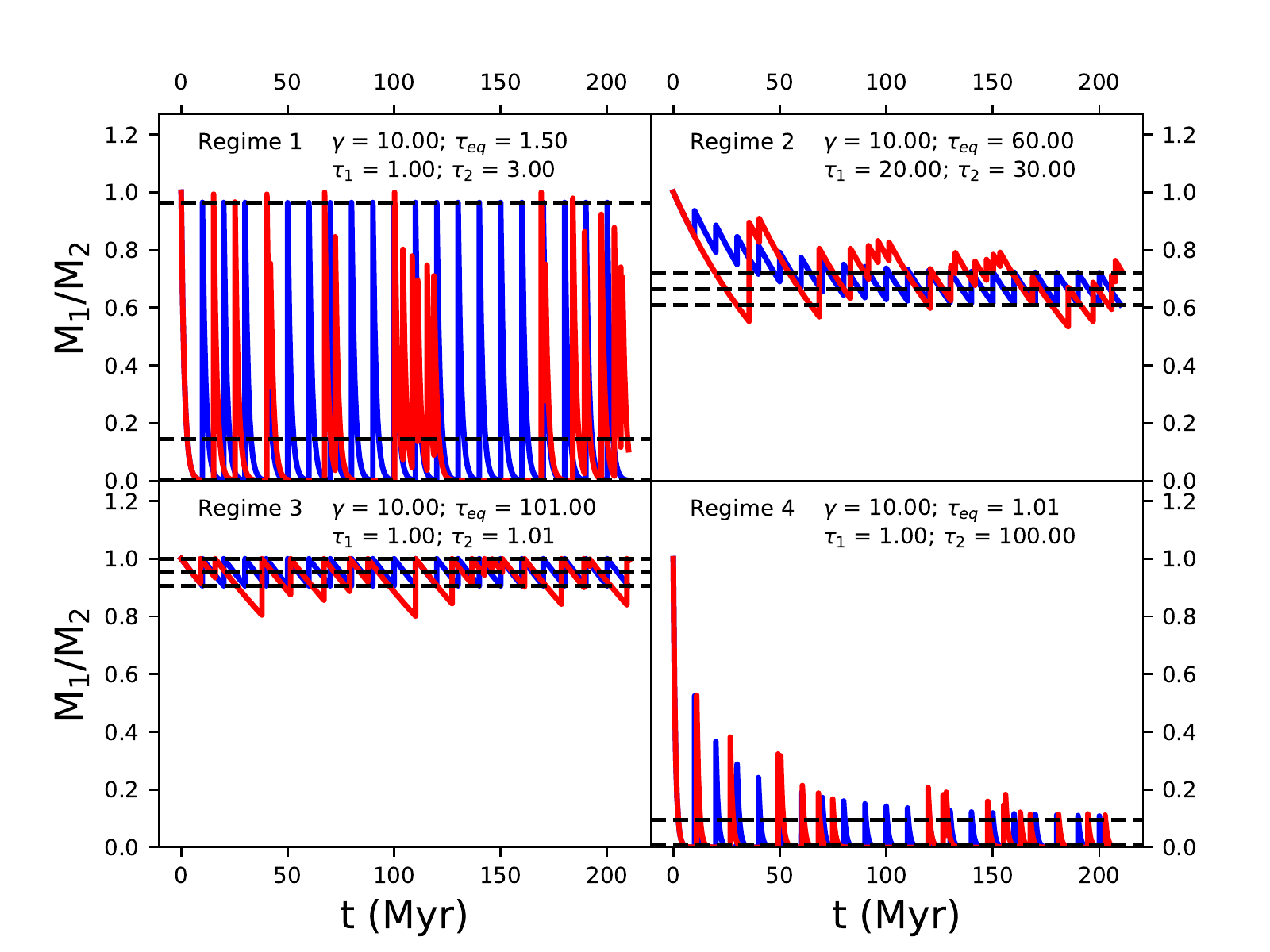}
    \caption{Examples of the behaviour of the four regimes explored in this work when $\tau_\text{eq} > 0$. The production ratio $P$ is taken to be $1$. The blue lines are the evolution for constant delta $\delta_c$, and the black, dashed lines correspond to the maximum, average and minimum values given by Equations (\ref{eq:avgRatio}) and (\ref{eq:maxMinRatio}), while the red lines are the evolution when $\delta$ is a random variable. In the figure annotation, $\gamma$ represents the time between the formation of two enrichment sources progenitor instead of the time between two actual successive enrichment events, exactly as defined in Paper I. As in that work, we find that $\gamma = \langle \delta \rangle$. The larger uncertainty of the stochastic case relatively to the $\delta_c$ case is readily apparent for all of the regimes.}
    \label{fig:FourRegimesStochastic}
\end{figure*}

Equation (\ref{eq:avgRatio}) is remarkably similar to that derived in \citet{Lugaro2018} for a SLR/stable isotopic ratio, with the main difference being that the mean life of the radioactive isotope $\tau$ is now substituted by the mean life of the ratio of the radioactive isotopes $\tau_\text{eq}$, and that now the multiplying exponentials do not cancel out\footnote{When considering only one radioactive isotope at the numerator, there is no exponential with $\tau_2$ at the numerator, and $\tau_\text{eq} = \tau_1$, leaving just $\tau/\delta_c$.}. The relative variation, that is, the difference between the maximum and minimum value divided by the average, is otherwise identical to the case of the SLR/stable isotopic ratio, provided we substitute the SLR mean life with the equivalent mean life. This means that, qualitatively, we can expect the uncertainty of the ratio between two radioactive isotopes to behave like that of a single radioactive isotope with mean life given by $\tau_\text{eq}$. However, the fact that the average value contains three non-vanishing exponentials means that, depending on the relative values of $\delta_c$, $\tau_1$, $\tau_2$, and $\tau_\text{eq}$, we face four qualitatively distinct regimes for the evolution of the ratio itself. These regimes are exemplified in Figure~\ref{fig:FourRegimesStochastic} and explained below.

\subsection{Regime 1; $\delta_c \gg \tau_\text{eq}, \tau_1, \tau_2$}

We study first the regime where $\delta_c \gg \tau_\text{eq}, \tau_1, \tau_2$. In this case, represented by the example on the top-left panel of Figure~\ref{fig:FourRegimesStochastic}, the average abundance ratio is
\begin{equation}
        \mu =
    \begin{cases}
        \frac{\tau_\text{eq}}{\delta_c}\,P& \text{if  } \tau_\text{eq} > 0,\\
        \frac{|\tau_\text{eq}|}{\delta_c}\,P\,e^{\delta_c/|\tau_\text{eq}|}& \text{if  } \tau_\text{eq} < 0.
    \end{cases}
    \label{eq:avgReg1}
\end{equation}
Given that the ratio $\tau_\text{eq}/\delta_c$ is small, we expect an average value much lower than the production ratio $P$ when $\tau_\text{eq} > 0$. For a case where $\tau_\text{eq} < 0$, we have an exponential term of $\delta_c/|\tau_\text{eq}|$, which will instead yield an average value much larger than $P$. In addition, the ratio will vary between the production ratio $P$ and 0 (or $P$ and $P\exp(\delta_c/|\tau_\text{eq}|)$) for the case of positive (negative) $\tau_\text{eq}$.

The intuitive understanding of this regime is that the time between enrichment events is longer than what it takes for both radioactive isotopes and their ratio to decay, which prevents any memory build-up and results in a very large relative uncertainty.

\subsection{Regime 2; $\delta_c \ll \tau_\text{eq}, \tau_1, \tau_2$}

In this regime, $\delta_c \ll \tau_\text{eq}, \tau_1, \tau_2$. This case, represented in the top-right panel of Figure~\ref{fig:FourRegimesStochastic}, has an equilibrium average value of
\begin{equation}
    \mu = P\,\frac{\tau_1}{\tau_2}.
    \label{eq:avgReg2}
\end{equation}

The evolution of the ratio of radioactive isotopes is marked by relatively frequent events, and the time between them is shorter than the mean life of any of the isotopes. This means that the abundance of both isotopes retains the memory of the previous events and the ratio drifts from the production ratio $P$ to oscillate around the equilibrium average with a low relative uncertainty, behaving in a similar fashion to the case of large $\tau/\delta_c$ studied in \citet{Lugaro2018}.

\subsection{Regime 3; $\delta_c \ll \tau_\text{eq}$ and $\delta_c \gg \tau_1, \tau_2$}

In this regime, $\delta_c \ll \tau_\text{eq}$ and $\delta_c \gg \tau_1, \tau_2$. This case, represented in the bottom-left panel of Figure~\ref{fig:FourRegimesStochastic} has an equilibrium average value of
\begin{equation}
    \mu = P.
    \label{eq:avgReg3}
\end{equation}

Although the value for the average in this case can be recovered from the formula of Regime 2 by using $\tau_1 \approx \tau_2$, we set this case apart because it represents the specific situation when the equivalent mean life is much larger than $\delta_c$, while the individual mean lives of each isotope are not. This regime only arises when the difference between the mean lives is small enough to make $\tau_\text{eq}$ orders of magnitude larger than them (see Eq.~\ref{eq:tauEq}). Given the short mean life of the individual SLR, it is likely that each SLR carries information from the last event only (see Paper I, Fig. 9 and related discussion). At the same time, the variation on the value of the ratio is relatively small because the equivalent mean life is too long for the ratio to change significantly before the next enriching event.

\subsection{Regime 4; $\delta_c \gg \tau_\text{eq}, \tau_1; \delta_c \ll \tau_2$}

In this regime $\delta_c \gg \tau_\text{eq}, \tau_1$, but $\delta_c \ll \tau_2$. The average value in this case, shown in the bottom-right panel of Figure~\ref{fig:FourRegimesStochastic}, is
\begin{equation}
        \mu =
    \begin{cases}
        \frac{\tau_\text{eq}}{\tau_2}\,P& \text{if  } \tau_\text{eq} > 0,\\
        \frac{|\tau_\text{eq}|\tau_1}{\delta_c^2} e^{\delta_c/|\tau_\text{eq}|}\,P& \text{if  } \tau_\text{eq} < 0.
    \end{cases}
    \label{eq:avgReg4}
\end{equation}
Although the evolution resembles that of the first regime when $\tau_\text{eq} > 0$, the maximum value attained by the ratio of the radioactive isotopes in the equilibrium becomes much lower than $P$. This is because, although the evolution of $M_1$ does not retain the memory of the previous events, the evolution of $M_2$ does. We note that in this regime $\tau_\text{eq} \approx \min(\tau_1, \tau_2)$.

\section{The case of variable $\delta$} \label{sec:stochasticDelta}

The cases studied in the previous section for a constant $\delta$ provide an intuition of how the ratio of two radioactive isotopes can behave in general. However, this simple approach produces deceptively small uncertainties relative to the more realistic scenario of variable $\delta$. This situation was explored already in Paper I for the case of the evolution of a single radioactive isotope, and it is illustrated here in Figure~\ref{fig:FourRegimesStochastic} also for the case of the ratio of two radioactive isotopes. To extend towards a better representation of SLR abundance variations in the ISM, we turn to a Monte Carlo approach where the enriching rate is stochastic, as in Paper I.

The set-up for the Monte Carlo experiments is the same as in Paper I. A total of 1000 runs are calculated for 15 Gyr each. For each run, the progenitors of the enriching events are generated with a constant time interval of $\gamma$. The time between the birth of the progenitor and the associated enriching event is sampled from a source-specific delay-time distribution (DTD). The enriching times are sorted and the random $\delta$ calculated from their consecutive differences (see Figure~2 of Paper I). Because the value for $\langle\delta\rangle$ is approximately that of $\gamma$, we use the terms interchangeably in this work.

The DTD used here have an equal probability between given initial and final times, and are the same as the ``box'' DTD of Paper I. We have omitted the ``power law'' DTD because, as concluded in Paper I, the actual $\delta$ distribution is approximately the same for both kinds of DTD for equal initial and final times. As in Paper I, we refer to the uniform distribution between 3 and 50 Myr, 50 Myr and 1 Gyr, and 50 Myr and 10 Gyr as the ``short'', ``medium'', and ``long'' box DTD, respectively. Each of these boxes can be associated with a different kind of progenitor for the enriching event, as described in Paper I.

Because in the synchronous case both radioisotopes are generated in the same events, the ratios are computed at each timestep for the same run. To explore the different regimes, each of the 1000 runs is repeated using different $\tau$. We consider 1000 runs to be enough for the same reasons as in Paper I: different temporal points of different runs are statistically independent and can, therefore, be considered as different experiments for the purposes of statistical derivation. For this reason, we stack together all the values between 10 and 14 Gyr to represent the final distribution of $M_1/M_2$. All the cases studied here have $\tau_1 < \tau_2$. This particular choice is arbitrary, however, cases with $\tau_1 > \tau_2$ result in positive exponential behavior, and the abundance ratio is no longer bounded and can diverge towards infinity, which complicates the analysis without adding any meaning,

\begin{figure*}
    \includegraphics[width=7.0in]{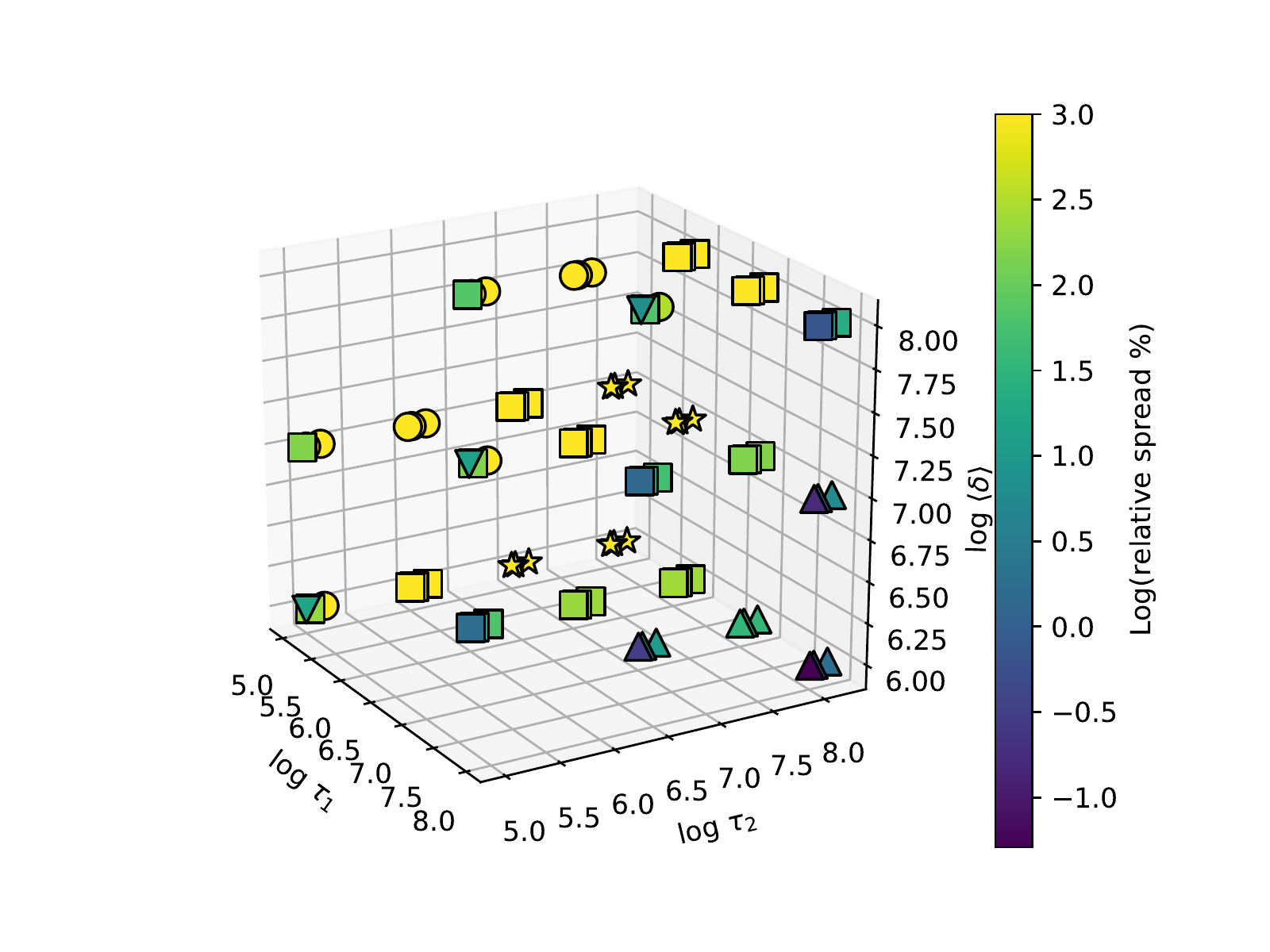}
    \caption{Dependence of the relative spread around the median on $\tau_1$, $\tau_2$ and $\langle\delta\rangle$. The 4 different regimes illustrated in Figure~\ref{fig:FourRegimesStochastic} cluster with different regions in this plot. Regimes 1 (circles) and 4 (stars) are located in the upper- and lower-left-far corner, respectively, with logarithmic relative spread values above 2 dex (100\%). Regime 2 (triangles) is located in the lower-right-far corner, with logarithmic relative spread values between 0 (1\%) and 1.5 dex (32\%), and Regime 3 (inverted triangles) is located on the diagonal contained in the $\tau_1 \approx \tau_2$ plane, with logarithmic relative spread lower than 1 dex. 
    Cases with the same $\tau_\text{eq}$ correspond to vertical lines with constant $\tau_1$ and $\tau_2$. Squares represent combinations that do not fall neatly into any regime and often correspond to a transition between two regimes.}
    \label{fig:3Dspread}
\end{figure*}

In Figure~\ref{fig:3Dspread} we show the relative uncertainty (68.2\% of the distribution around the median of the ratio) resulting from the Monte Carlo experiments when varying $\tau_1$, $\tau_2$, and $\gamma$. As the figure shows, Regimes 1 and 4 
have extremely large relative uncertainties, mainly due to $M_1$ not building up sufficient memory. Therefore, these regimes can only be treated as additions of individual events, using statistical methods different from that used here. This is similar to the case of Regime II of Paper I (all the regimes of Paper I and their connections to the present regimes will be described in more detail in Section~\ref{sec:PaperI}). Therefore, from now on we will focus on the cases where $\tau_\text{eq} \gtrsim 3\langle\delta\rangle$, which excludes Regimes 1 and 4. The exception is Regime 3, where although neither $M_1$ nor $M_2$ build up enough memory from previous events, the slow decaying property of their ratio results in a stable value with low relative uncertainty. This makes Regime 3 an interesting case where the uncertainty in the ratio of two SLR is as low or lower than in Regime 2, with a large percentage of the ratio containing only the abundances from the last event.

\begin{deluxetable*}{lcccccc}
\tablewidth{0pc}
\tablecaption{Median values and 68\% confidence interval for cases belonging to Regime 2 and lower limits encompassing the 84\% of the distribution for cases belonging to Regime 3, from the Monte Carlo experiment (for $P = 1$), for different values of $\gamma$, $\tau_1/\gamma$, $\tau_2/\gamma$ and $\tau_\text{eq}/\gamma$ for $\tau_\text{eq}/\gamma > 3$. The results from the large box are identical to those from the medium box DTD. A dash in the Regime column means that the specific case does not neatly fall neatly into one of the regimes. These cases typically fall between Regime 1 or 4 and Regime 3. \label{tab:MCTableVals}}
\tablehead{
$\gamma$ [Myr] & $\tau_1/\gamma$ & $\tau_2/\gamma$ & $\tau_\text{eq}/\gamma$ & Small box & Large box & Regime
}
\startdata
$1.00$ & $0.10$ & $0.10$ & $10.10$ & $> 0.83$ & $> 0.83$ & $3$\\
$1.00$ & $1.00$ & $1.01$ & $101.00$ & $> 0.98$ & $> 0.98$ & $3$\\
$1.00$ & $1.00$ & $1.10$ & $11.00$ & $> 0.80$ & $> 0.80$ & $3$\\
$1.00$ & $1.00$ & $1.50$ & $3.00$ & $0.63_{-0.20}^{+0.13}$ & $0.63_{-0.20}^{+0.13}$ & $-$\\
$1.00$ & $10.00$ & $10.10$ & $1010.00$ & $0.99_{-0.00}^{+0.00}$ & $0.99_{-0.00}^{+0.00}$ & $2$\\
$1.00$ & $10.00$ & $11.00$ & $110.00$ & $0.91_{-0.01}^{+0.01}$ & $0.91_{-0.02}^{+0.01}$ & $2$\\
$1.00$ & $10.00$ & $15.00$ & $30.00$ & $0.66_{-0.04}^{+0.03}$ & $0.66_{-0.04}^{+0.04}$ & $2$\\
$1.00$ & $10.00$ & $101.00$ & $11.10$ & $0.10_{-0.02}^{+0.02}$ & $0.10_{-0.02}^{+0.02}$ & $2$\\
$1.00$ & $10.00$ & $110.00$ & $11.00$ & $0.09_{-0.01}^{+0.02}$ & $0.09_{-0.02}^{+0.02}$ & $2$\\
$1.00$ & $10.00$ & $150.00$ & $10.71$ & $0.07_{-0.01}^{+0.01}$ & $0.07_{-0.01}^{+0.01}$ & $2$\\
$1.00$ & $100.00$ & $101.00$ & $10100.00$ & $0.99_{-0.00}^{+0.00}$ & $0.99_{-0.00}^{+0.00}$ & $2$\\
$1.00$ & $100.00$ & $110.00$ & $1100.00$ & $0.91_{-0.00}^{+0.00}$ & $0.91_{-0.00}^{+0.00}$ & $2$\\
$1.00$ & $100.00$ & $150.00$ & $300.00$ & $0.67_{-0.01}^{+0.01}$ & $0.67_{-0.01}^{+0.01}$ & $2$\\
$10.00$ & $0.10$ & $0.10$ & $10.10$ & $> 0.86$ & $> 0.83$ & $3$\\
$10.00$ & $1.00$ & $1.01$ & $101.00$ & $> 0.98$ & $> 0.98$ & $3$\\
$10.00$ & $1.00$ & $1.10$ & $11.00$ & $> 0.83$ & $> 0.80$ & $3$\\
$10.00$ & $1.00$ & $1.50$ & $3.00$ & $0.64_{-0.17}^{+0.12}$ & $0.63_{-0.20}^{+0.13}$ & $-$\\
$10.00$ & $10.00$ & $10.10$ & $1010.00$ & $0.99_{-0.00}^{+0.00}$ & $0.99_{-0.00}^{+0.00}$ & $2$\\
$10.00$ & $10.00$ & $11.00$ & $110.00$ & $0.91_{-0.01}^{+0.01}$ & $0.91_{-0.01}^{+0.01}$ & $2$\\
$10.00$ & $10.00$ & $15.00$ & $30.00$ & $0.67_{-0.02}^{+0.02}$ & $0.66_{-0.04}^{+0.04}$ & $2$\\
$100.00$ & $0.10$ & $0.10$ & $10.10$ & $0.95_{-0.03}^{+0.03}$ & $> 0.83$ & $3$\\
$100.00$ & $1.00$ & $1.01$ & $101.00$ & $0.99_{-0.00}^{+0.00}$ & $> 0.98$ & $3$\\
$100.00$ & $1.00$ & $1.10$ & $11.00$ & $0.90_{-0.03}^{+0.03}$ & $> 0.80$ & $3$\\
$100.00$ & $1.00$ & $1.50$ & $3.00$ & $0.65_{-0.08}^{+0.08}$ & $0.63_{-0.20}^{+0.13}$ & $-$\\
\enddata
\end{deluxetable*}

The uncertainties from the Monte Carlo calculations are presented in Table~\ref{tab:MCTableVals} for $\tau_\text{eq} > 0$ and $\tau_\text{eq}/\gamma > 3$. When the distribution is approximately symmetric (Regime 2), both an upper and lower value are given, when the distribution piles-up on $P$ (Regime 3), a lower limit for the ratio is given instead. Table \ref{tab:MCTableVals} allows us to calculate uncertainties for ratios of SLR due to temporal stochasticity of enrichment events. For any isotopic ratio, we can select the proper $\gamma$, which depends on the source, the best suited $\tau_1/\gamma$ and $\tau_2/\gamma$, and whether a short box (i.e., if the source are core collapse supernovae) or a long box (i.e., if the source are asymptotic giant branch stars or neutron star mergers) describes the source. Afterwards, the corresponding numbers in Column 5 or 6 should be multiplied by the production ratio of the SLR ratio. If there is no exact match to the numbers shown in Table~\ref{tab:MCTableVals}, then Equations~(\ref{eq:approxAvg}) and (\ref{eq:approxSigma2}) or (\ref{eq:approxSigma3}) described below in Section \ref{sec:analyticalApproach}) can be used instead. In Sections~\ref{sec:regime2} and \ref{sec:regime3}, we describe in more detail the differences between the constant and random $\delta$ cases in relation to Regimes 2 and 3, respectively.

\subsection{Connections and similarities with the regimes defined in Paper I}
\label{sec:PaperI}

In Paper I we analysed a single SLR and found 3 different regimes can be applied depending on the relation between $\tau$ and $\gamma$. Here we report a brief description of them and and how they connect with the regimes in this work. For sake of clarity, the 3 regimes from Paper I are marked in Roman numerals, while Arabic numerals refer to the 4 regimes considered here.

Regime I refers to $\tau/\gamma > 2$ and it is similar to Regimes 2 or 3, in that statistics can be calculated because the spread is not much larger than the median value. Regime I is associated with the calculation of the isolation time, T$_\text{iso}$, because in this case the ISM will contain an equilibrium value from where there can be an isolation period before the ESS abundances. In the present work, Regime 2 is that associated with the calculation of T$_\text{iso}$.

Regime III is a case that covers the region of $\tau/\gamma < 0.3$. In this Regime, there is a large probability that the ISM abundance that decayed into the ESS abundance originated from a single event. Therefore, this Regime is associated with the calculation of the time since the last event, T$_\text{LE}$. Regime 3 of this work is related to Regime III of Paper I in that both carry most likely abundances from only the last event before the formation of the Solar System. The difference is that, while Regime III allows us to calculate T$_\text{LE}$, Regime 3 allows us to also narrowly determine the production ratio of the last event.

Regime II falls between two well-defined cases described above. This regime has $0.3 < \tau/\gamma < 2$, which does not allow for meaningful statistics nor for a clean definition of a last event to which the ISM abundance can be solely or mostly attributed. This Regime does not correspond to any of the regimes in this work, and it may be similar to the region between Regime 2 and Regimes 1 and 4.

\subsection{Analytical approach}\label{sec:analyticalApproach}

We also investigated the possibility to calculate the uncertainties using an analytical approach instead of the full Monte Carlo simulations. The aim is to provide a better understanding of the regimes and their uncertainties, as well as give an alternative to calculate approximate numbers without the need of a simulation. To do that, we use the expression for the average given by
\begin{equation}
    \mu \approx P\,\frac{\tau_\text{eq}}{\langle\delta\rangle}\frac{1 - \langle e^{-\delta/\tau_2}\rangle}{1 - \langle e^{-\delta/\tau_1}\rangle} \left(1 - \langle e^{-\delta/\tau_\text{eq}} \rangle \right),
    \label{eq:approxAvg}
\end{equation}
derived in Appendix \ref{sec:mathDevelopment}, and for the relative standard deviation we use
\begin{equation}
    \frac{\sigma}{\mu} \approx F\sqrt{\frac{\langle\delta\rangle}{2\tau_\text{eq}}\frac{1 - \langle e^{-2\delta/\tau_\text{eq}} \rangle}{\left(1 - \langle e^{-\delta/\tau_\text{eq}}\rangle\right)^2} - 1},
    \label{eq:approxSigma2}
\end{equation}
where $F$ is a correction factor applied to Equation \ref{eq:approxSigma1} and is defined by
\begin{equation}
    F = K\left[1 + \log_{10}\left(\frac{\min(\tau_1, \tau_2)}{\langle\delta\rangle}\right)^2\right],
    \label{eq:eqForF}
\end{equation}
where $K = 1$ unless $\min(\tau_1, \tau_2)$ is larger than the span of the DTD, in which case $K = 0.5$. In cases where $\min(\tau_1, \tau_2) < \langle\delta\rangle$, then $F = K$.

This factor $F$ was derived from the Monte Carlo experiments and corrects some of the approximations made in the derivation of \ref{eq:approxSigma1} in Appendix \ref{sec:mathDevelopment}. With this correction factor, Equation (\ref{eq:approxSigma2}) becomes an accurate estimation to the results of the Monte Carlo experiment.

If the full distribution of $\delta$ is unknown, a further approximation to Equation (\ref{eq:approxSigma2}) can be used instead, rendering
\begin{equation}
    \frac{\sigma}{\mu} \approx F \sqrt{\frac{1}{6}\frac{\langle\delta\rangle^3 + 3\sigma_\delta^2\langle\delta\rangle}{\tau_\text{eq}^2\langle\delta\rangle - \tau_\text{eq}\sigma_\delta^2}},
    \label{eq:approxSigma3}
\end{equation}
with the advantage that only $\langle\delta\rangle$ and $\sigma_\delta$ (the standard deviation of the delta distribution) have to be known. This formula is much easier to calculate because no sampling of the $\delta$ distribution is needed.

\begin{figure}
    \includegraphics[width=0.5\textwidth]{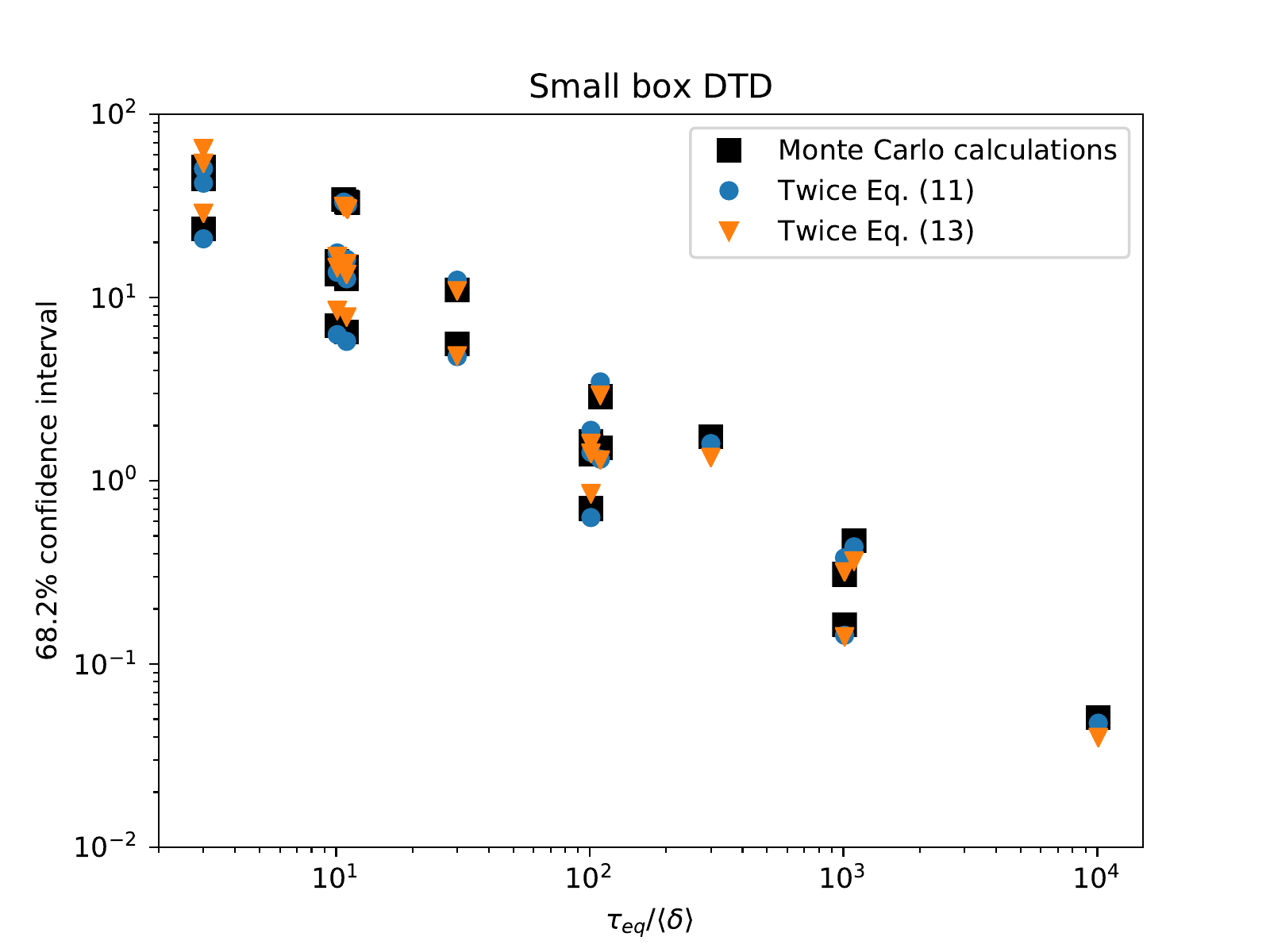}
    \caption{Prediction of Equations~(\ref{eq:approxSigma2}) and (\ref{eq:approxSigma3}) of the relative to the median 68.2\% confidence interval calculated from the Monte Carlo experiments (black large squares) for the small box DTD. The equations themselves calculate just the 34.1\% interval, which is why twice their value is used.}
    \label{fig:SmallBoxSpread}
\end{figure}

The validity of Equations (\ref{eq:approxSigma2}) and (\ref{eq:approxSigma3}) can be tested by comparing it to the 68.2\% (1$\sigma$) confidence interval calculated from the Monte Carlo experiments. This comparison is presented in Figure~\ref{fig:SmallBoxSpread}. In the worst case, with the small box DTD, the relative difference between the analytical approximations and the results from the numerical experiments is just above 25\%. These are valid for calculations related to Regime 2. For Regime 3, instead, as seen in Table \ref{tab:MCTableVals}, the average remains very close to $P$ introduces asymmetry in the distribution. In this case, the theoretical $\sigma$ is an average of the lower and upper $1\sigma$ threshold. When this $\sigma$ is such that $\mu + \sigma > P$, it is better to calculate a lower limit for the distribution with $P - 2\sigma$, because in this cases the distribution piles up at $P$, making any value between $P$ and $\mu$ functionally equiprobable.

\subsection{Regime 2; $\delta \ll \tau_\text{eq}, \tau_1, \tau_2$}
\label{sec:regime2}

In this case the abundances of both SLR nuclei retain significant memory from past events. The average of their ratio, according to Equation~(\ref{eq:approxAvg}), is the same as the constant case for the same regime, given by Equation~(\ref{eq:avgReg2}). When comparing the uncertainties, however, there is a significant difference between the constant and the stochastic case. As a first order approximation, and taking $\sigma_\delta \approx \langle\delta\rangle$ (see Table 2 of Paper I), we can write Equation (\ref{eq:approxSigma3}) as
\begin{equation}
    \frac{2\sigma}{\mu} \approx 2F \frac{\langle\delta\rangle}{\tau_\text{eq}},
\end{equation}
which, when substituting $\langle\delta\rangle$ by $\delta_c$, dividing Eqation~(\ref{eq:eqForF}) by Equation~(\ref{eq:maxMinRatio}) reveals that the stochastic case has a larger uncertainty relative to the constant case by a factor of 2F. This factor can be shown to be in the range $2F \in [2.5, 35]$ when considering $\tau_\text{eq}/\langle\delta\rangle \in [3, 10^4]$ by using Equation~(\ref{eq:eqForF}) with $K = 1$ and taking $\min(\tau_1, \tau_2) = \tau_\text{eq}$. 
Therefore, the time-stochastic nature of enrichment events can increase the uncertainty by more than an order of magnitude in this regime.
The uncertainty on the ratio of two SLR in Regime 2 is still relatively low. For example, for the Large Box, with $\tau_1 = 10$ Myr, $\tau_2 = 15$ Myr and $\gamma = 1$ Myr, Table~\ref{tab:MCTableVals} has a relative uncertainty of $12\%$. For a similar example with $\tau = 10$ Myr and $\gamma = 1$ Myr in Table~3 of Paper I, the relative uncertainty is $45\%$ for the Large Box. Even if we take the case of $\tau = 31.6$ Myr and $\gamma = 1$ Myr, we still have a relative uncertainty of $25\%$ for the SLR/stable isotopic ratio.

\subsection{Regime 3; $\delta \ll \tau_\text{eq}$, $\delta \gg \tau_1, \tau_2$}
\label{sec:regime3}

As discussed in the constant $\delta_c$ scenario, this regime shows a low variation around the average $P$ while retaining no memory of previous events. The difference between the constant and stochastic case is similar to that in Regime 2 (because Equation~\ref{eq:approxSigma3} depends only on $\langle\delta\rangle$ and $\tau_\text{eq}$), that is, a factor of $2F$. The factor $F = K$ is a constant here (when $\min(\tau_1, \tau_2) < \langle\delta\rangle$), equal either to 0.5 or to 1, which means that the relation between the uncertainties in the constant and stochastic cases is of a factor of two at most. Additionally, the stochastic case results in a non-symmetric distribution around the median. The reason is that the ratio is always bounded between 0 and the production factor $P$ (when $\tau_2 > \tau_1$): when the enriching events are more frequent than average, the ratio will remain at $P$, while when the enriching events are less frequent than average the ratio decays away from this average. In any case, the characteristic of  Regime 3 is that the average ratio remains always very close to $P$. 

\section{Discussion} \label{sec:discussion}

We apply our general theoretical approach to specific ratios of two SLRs that are either in Regime 2 or Regime 3. Starting from Table~2 of \citet{Lugaro2018}, which lists all the SLRs known to have been present in the ESS, we select SLRs with potentially the same origin (for the synchronous scenario) and with mean lives close enough such that the $\tau_\text{eq}$ of their ratio is potentially larger than the probable $\gamma$ of their source. We find four cases of such ratios of isotopes and present them in Table~\ref{tab:specificValuesSync}, along with the specific Monte Carlo (MC) experiments that reproduce the conditions under which they evolve in the Galaxy, assuming a production ratio $P = 1$. This table categorizes the regime of the selected SLR ratios, realizes the difference with regards to the uncertainties between considering the single SLR/stable (or long-lived) reference isotope ratio (Columns SLR$_{1}$ and SLR$_{2}$), and quantifies the ratio of the two SLRs (Column SLR$_{1}$/SLR$_{2}$). In general, the uncertainties significantly decrease when considering ratios of SLRs with similar mean lives, relative to considering their ratio to a stable or long-lived isotope (compare the last column of MC values to the other two columns of MC values). It is worth mentioning that in this comparison we are supposing that the stable isotope carries no uncertainty at all from GCE processes, which by itself can be a factor of up to $5.7/1.6 = 3.6$ \citep{Cote2019}. In addition, the predicted ISM abundances are much closer to the production ratios when considering ratios between two SLRs.

Table~\ref{tab:ESSvaluesTiso} shows the subsequent calculations of the isolation time, $T_\text{iso}$ (in roman), and the time since the last event, $T_\text{LE}$ (in italics), for the selected isotopic ratios for which the ESS ratio is available. These correspond to only three out of the four ratios discussed in Table~\ref{tab:specificValuesSync}. We excluded $^{97}$Tc/$^{98}$Tc because only upper limits are available for the corresponding radioactive-to-stable ratios, which means it is not possible to derive any ESS value for their ratio. The other ESS ratios are calculated using the values for the radioactive-to-stable ratios reported in Table~2 of \citet{Lugaro2018} and the solar abundances of the reference isotopes from \citet{Lodders2010} \citep[see also][]{Cote2020}. Furthermore, the selected values for $\gamma$ were limited to those most likely to occur in the Milky Way for the corresponding production sites. 

\begin{deluxetable*}{lcccccccccc}
\tablewidth{0pc}
\tablecaption{Regimes and values of the ratios from the Monte Carlo (MC) experiments applied to the specific cases of ratios between two SLRs (Column SRL$_{1}$/SRL$_{2}$) and between the SLRs and their corresponding stable or long-lived reference isotopes (Columns SRL$_1$/stable and SRL$_2$/stable, see the main text for the list of reference isotopes). Production ratios are always 1. Also indicated are $\tau_1$, $\tau_2$, $\tau_\text{eq}$ and the adopted $\gamma$, all in Myr. The values of $\gamma$ are selected such that it is possible to remain within Regimes 2 or 3, for which cases we can model the uncertainties. The Roman numerals correspond to the regimes of Paper I (SRL$_{1,2}$/stable), while the Arabic numerals correspond to the regimes described in this work. The hyphen symbols correspond to cases that do not fit neatly in any of the regimes. \label{tab:specificValuesSync}}
\tablehead{
& \multirow{2}{*}{$\tau_1$} & \multirow{2}{*}{$\tau_2$} & \multirow{2}{*}{$\tau_\text{eq}$} & \multirow{2}{*}{$\gamma$} & \multicolumn{2}{c}{SLR$_1$/stable} & \multicolumn{2}{c}{SLR$_2$/stable} & \multicolumn{2}{c}{SLR$_1$/SLR$_2$}\\
& & & & & Regime & MC values & Regime & MC values & Regime & MC values
}
\startdata
\multirow{6}{*}{$^{247}$Cm/$^{129}$I} & \multirow{6}{*}{22.5} & \multirow{6}{*}{22.6} & \multirow{6}{*}{5085}
& 1 & I & $22.37^{+3.45}_{-3.22}$ & I & $22.47^{+3.46}_{-3.23}$ & 2 & $1.00^{+0.00}_{-0.00}$\\
& & & & 3.16 & I & $7.01^{+1.98}_{-1.77}$ & I & $7.04^{+1.99}_{-1.77}$ & 2 & $1.00^{+0.00}_{-0.00}$\\
& & & & 10 & II & $< 3.29$ & II & $< 3.30$ & 3 & $1.00^{+0.00}_{-0.00}$\\
& & & & 31.6 & II & $< 1.29$ & II & $< 1.29$ & 3 & $> 0.99$\\
& & & & 100 & III & $< 0.54$ & III & $< 0.54$ & 3 & $> 0.96$\\
& & & & 316 & III & $< 0.08$ & III & $< 0.08$ & 3 & $> 0.89$\\
\hline
\multirow{4}{*}{$^{107}$Pd/$^{182}$Hf} & \multirow{4}{*}{9.4} & \multirow{4}{*}{12.8} & \multirow{4}{*}{35.4}
& 1 & I & $9.28^{+2.27}_{-2.05}$ & I & $12.68^{+2.63}_{-2.41}$ & 2 & $0.73^{+0.03}_{-0.04}$\\
& & & & 3.16 & I & $2.86^{+1.32}_{-1.10}$ & I & $3.94^{+1.53}_{-1.31}$ & 2 & $0.73^{+0.06}_{-0.08}$\\
& & & & 10 & II & $< 1.62$ & II & $< 2.06$ & 3 & $0.70^{+0.12}_{-0.19}$\\
& & & & 31.6 & III & $< 0.69$ & III & $< 0.86$ & - & $0.50^{+0.30}_{-0.32}$\\
\hline
\multirow{4}{*}{$^{53}$Mn/$^{97}$Tc} & \multirow{4}{*}{5.4} & \multirow{4}{*}{5.94} & \multirow{4}{*}{59.4}
& 1 & I & $5.29^{+1.75}_{-1.53}$ & I & $5.83^{+1.83}_{-1.61}$ & 2 & $0.91^{+0.02}_{-0.02}$\\
& & & & 3.16 & II & $< 2.63$ & II & $< 2.84$ & 3 & $0.90^{+0.03}_{-0.05}$\\
& & & & 10 & II & $< 1.04$ & II & $< 1.12$ & 3 & $0.86^{+0.08}_{-0.15}$\\
& & & & 31.6 & III & $< 0.41$ & III & $< 0.45$ & - & $0.68^{+0.22}_{-0.31}$\\
\hline
\multirow{5}{*}{$^{97}$Tc/$^{98}$Tc} & \multirow{5}{*}{5.94} & \multirow{5}{*}{6.1} & \multirow{5}{*}{226}
& 1 & I & $5.83^{+1.83}_{-1.61}$ & I & $5.99^{+1.85}_{-1.63}$ & 2 & $0.97^{+0.00}_{-0.01}$\\
& & & & 3.16 & II & $< 2.84$ & II & $< 2.91$ & 3 & $0.97^{+0.01}_{-0.02}$\\
& & & & 10 & II & $< 1.12$ & II & $< 1.14$ & 3 & $0.96^{+0.02}_{-0.05}$\\
& & & & 31.6 & III & $< 0.45$ & III & $< 0.47$ & 3 & $0.90^{+0.07}_{-0.13}$\\
& & & & 100 & III & $< 0.05$ & III & $< 0.06$ & - & $0.73^{+0.19}_{-0.29}$\\
\enddata
\end{deluxetable*}

\begin{deluxetable*}{lccccccccc}
\tablewidth{0pc}
\tablecaption{Timescales derived by decaying the reported ISM ratios to the ESS ratios in Column 2 for a subset of ratios and $\gamma$ values considered in Table~\ref{tab:specificValuesSync} to represent possible realistic values in the Galaxy for the corresponding production event. Time and $\tau_\text{eq}$ are in Myr. The ISM SLR$_{1,2}$/stable ratios in roman are calculated using the steady-state formula from \citet{Cote2019} and K=2.3. These are cases within Regime I and can provide T$_\text{iso}$ (also in roman). The ISM SLR$_{1,2}$/stable ratios in italics are calculated instead using the last-event formula, i.e., Eqs.~3 (with K=2.3) and 4 (with K=1.2) of \citet{Cote2020} and the selected value of $\gamma=\delta$. These are cases within Regime III and can provide T$_\text{LE}$ (also in italics). The ISM SLR$_{1}$/SLR$_{2}$ values are calculated as $= P (\tau_1/\tau_2)$ (Eq.~\ref{eq:avgReg2}) for roman values and as $= P$ (Eq.~\ref{eq:avgReg3}) for italic values. The production ratios used in all the formulas are reported in the text in each subsection.
The ``back-decayed'' ratios are calculated by decaying back the ESS ratio by the average of T$_\text{iso}$, or T$_\text{LE}$, from both SLR$_{1,2}$/Stable ratios, except for the case of $^{53}$Mn/$^{97}$Tc, where only the times derived from $^{53}$Mn were used. Differences between the values in the last two columns highlight the problems discussed in the text.
\label{tab:ESSvaluesTiso}}
\tablehead{
& \multirow{2}{*}{ESS ratio} & \multirow{2}{*}{$\tau_\text{eq}$} & \multirow{2}{*}{$\gamma$} & \multicolumn{2}{c}{SLR$_1$/stable} & \multicolumn{2}{c}{SLR$_2$/stable} & \multicolumn{2}{c}{SLR$_1$/SLR$_2$}\\
& & & & ISM ratio & Time & ISM ratio & Time & ISM ratio & back-decayed ratio
}
\startdata
{$^{247}$Cm/$^{129}$I} & {$2.28 \times 10^{-3}$} & {5085}
& 316 & {\it $\mathit{9.63 \times 10^{-2}}$} & {\it 171} & {\it $\mathit{1.15 \times 10^{-1}}$ } & {\it 153} & {$\mathit{1.22 \times 10^{-2}}$($^a$)} & \textit{$\mathit{2.35 \times 10^{-3}}$}\\
\hline
\multirow{3}{*}{$^{107}$Pd/$^{182}$Hf} & \multirow{3}{*}{$4.25$} & \multirow{3}{*}{35.4}
& 1 & \multirow{2}{*}{$3.56 \times 10^{-4}$} & $16^{+2}_{-3}$ & \multirow{2}{*}{$5.20 \times 10^{-4}$} & $21^{+2}_{-3}$ & \multirow{2}{*}{2.41} & \multirow{2}{*}{7.17}\\
& & & 3.16 & & $16^{+4}_{-6}$ & & $21^{+4}_{-6}$ & & \\
& & & 31.6 & {\it $\mathit{1.20 \times 10^{-3}}$} & {\it 27} & {\it $\mathit{1.28 \times 10^{-3}}$} & {\it 32} & {\it 3.28} &  {\it 9.78} \\
\hline
\multirow{2}{*}{$^{53}$Mn/$^{97}$Tc} & \multirow{2}{*}{$> 1.70 \times 10^5$} & \multirow{2}{*}{59.4}
& 1 & $1.58 \times 10^{-4}$ & $17^{+2}_{-2}$ & $3.84 \times 10^{-5}$ & $> 7$ & $1.65 \times 10^6$ & $>2.26 \times 10^5$\\
& & & 31.6 & {\it $\mathit{9.23 \times 10^{-4}}$} & {\it 26} & {\it $\mathit{2.04 \times 10^{-4}}$} & {\it $\mathit{> 17}$} & {\it $\mathit{1.82 \times 10^6}$} & {\it $\mathit{> 2.63 \times 10^5}$} \\
\enddata
$^a$We calculated this possible \textit{r}-process production using the average $^{247}$Cm/$^{232}$Th ratio from \citet{Goriely2016} and assuming the solar ratio $^{127}$I/$^{232}$Th of $31$ from \citet{Asplund2009}. This is to avoid using $^{235}$U, which decays much faster than $^{232}$Th (with mean life of roughly 1 Gyr, instead of 20 Gyr) and would complicate the assumption that the produced  $^{127}$I/$^{235}$U was solar.
\end{deluxetable*}

\subsection{The ratio of the \textit{r}-process $^{247}$Cm and $^{129}$I}

These two isotopes are made by the \textit{rapid} neutron-capture (\textit{r}) process and typical estimates for the time interval at which \textit{r}-process nucleosynthetic events that are believed to enrich a parcel of gas in the Galaxy range between 200 and 500 Myr \citep{Hotokezaka15,Tsujimoto17,Bartos19,Cote2020}. Therefore, the case of $^{247}$Cm/$^{129}$I is the best example of Regime 3 since $\tau_\text{eq} = 5085$ Myr (Table \ref{tab:specificValuesSync}) is much larger than $\gamma$, while each $\tau$ ($\simeq$ 22.5 and 22.6 Myr, respectively) is much shorter than $\gamma$. The ratios to the long-lived or stable references isotopes, $^{247}$Cm/$^{235}$U and 
$^{129}$I/$^{127}$I, allow us to derive a T$_\text{LE}$, for example for the specific $\gamma$ value of 316 Myr considered in Table~\ref{tab:ESSvaluesTiso}, and derive typical production ratios of $1.35$ for $^{129}$I/$^{127}$I and $0.3$ for $^{247}$Cm/$^{235}$U. 
While our T$_\text{LE}$ values are not perfectly compatible with each other, the more detailed analysis shown by \citet{Cote2020} demonstrates that there is compatibility for T$_\text{LE}$ in the range between 100 - 200 Myr, depending on the exact choice of the K parameter \citep{Cote2019}, $\gamma$, and the production ratios. The short mean lives of $^{247}$Cm and $^{129}$I ensure that there is no memory from previous events, while the long $\tau_\text{eq}$ of $^{247}$Cm/$^{129}$I instead ensures that this ratio did not change significantly during T$_\text{LE}$ and has a high probability to be within the 10\% of the production ratio. Therefore, the production ratio of the last \textit{r}-process event that polluted the ESS material can be accurately determined directly from the ESS ratio. If we assume that the last event produced a $^{247}$Cm/$^{232}$Th ratio similar to the average predicted by \citet{Goriely2016}, and assume a solar ratio for $^{127}$I/$^{232}$Th, then we find an inconsistency between the numbers in the last two columns of Table~\ref{tab:ESSvaluesTiso}. The back-decayed value is more than five times lower than the assumed production ratio, which indicates a weaker production of the actinides from this last event, with respect to the production ratios that we are using here. The number in the last column represents therefore a unique constrain on the nature of the astrophysical sites of the \textit{r} process in Galaxy at the time of the formation of the Sun and needs to be compared directly to different possible astrophysical and nuclear models \citep{Cote2020}. 

\subsection{The ratio of the $s$-process $^{107}$Pd and $^{182}$Hf}

If T$_\text{LE}$ for the last \textit{r}-process event is larger than 100 Myr, as discussed in the previous section, the presence of these two SLRs in the ESS should primarily be attributed to the \textit{slow} neutron-capture (\textit{s}) process in asymptotic giant branch (AGB) stars, which are a much more frequent event due to the low mass of their progenitors, since their \textit{r}-process contribution would have decayed for a time of the order of 10 times their mean lives \citep{Lugaro2014}.
Experimental results on the SLRs $^{107}$Pd ($\tau$=9.8 Myr) and $^{182}$Hf ($\tau$=12.8 Myr) are reported with respect to the stable reference isotopes $^{108}$Pd and $^{180}$Hf, respectively. The ISM ratio reported in Table~\ref{tab:ESSvaluesTiso} are calculated using production ratios of 0.14, 0.15, and 3.28 for $^{107}$Pd/$^{108}$Pd, $^{182}$Hf/$^{180}$Hf, $^{107}$Pd/$^{182}$Hf, respectively, derived from the 3 M$_{\odot}$ model of \citet{Lugaro2014}.
For the short $\gamma$ values considered in Table~\ref{tab:ESSvaluesTiso} (1 and 3.16 Myr) the SLR$_{1,2}$/Stable ratios belong to Regime I and the SLR$_{1}$/SLR$_{2}$ ratio belong to Regime 2. Therefore, we can calculate the T$_\text{iso}$ from all the ratios. 
As shown in Table~\ref{tab:specificValuesSync}, the ratios relative to the stable reference isotopes suffer from larger uncertainties (40\% or 85\% depending on the $\gamma$, and supposing no uncertainty in the stable isotope abundance) compared to the ratio of the two SLRs (less than 20\%). However, when considering 
the actual ISM ratios, the uncertainties on the evaluation of T$_\text{iso}$ become comparable because these are relative uncertainties and the ratio of the two SLRs and the equivalent mean life have a much larger absolute value that the other two ratios. While the T$_\text{iso}$ values derived from the SLR$_{1,2}$/stable ratios are consistent with each other, the value calculated from SLR$_{1}$/SLR$_{2}$ would need to be much shorter. In the last column of Table~\ref{tab:ESSvaluesTiso} we report the back-decayed ratio, as the ISM ratio that is required to obtain a self-consistent solution. 

The discrepancy between the ISM and back-decayed values may be due to problems with the stellar production of these isotopes: a main caveat here to consider is that, while the $^{107}$Pd/$^{108}$Pd ratio produced by the \textit{s} process is relatively constant, since it only depends on the inverse of the ratio of the neutron-capture cross sections of the two isotopes, both the $^{182}$Hf/$^{180}$Hf and $^{107}$Pd/$^{182}$Hf production ratios can vary significantly between different AGB star sources. The $^{182}$Hf/$^{180}$Hf ratio is particularly sensitive to the stellar mass \citep{Lugaro2014}, due to the probability on the activation of the $^{181}$Hf branching point, which increases with the neutron density produced by the $^{22}$Ne($\alpha$,n)$^{25}$Mg neutron source reaction, which, in turn, increases with temperature and therefore stellar mass. The $^{107}$Pd/$^{182}$Hf involves two isotopes belonging to the mass region before ($^{107}$Pd) and after ($^{182}$Hf) the magic neutron number of 82 at Ba, La, and Ce. This means that this ratio will also be affected by the total number of neutrons released by the main neutron source $^{13}$C($\alpha$,n)$^{16}$O in AGB stars, which has a strong metallicity dependence \citep[see, e.g.,][]{Gallino1998,Cseh2018}. This means that a proper analysis of these $s$-process isotopes can only be carried out in the framework of a full GCE models, where the stellar yields are varied with mass and metallicity. This work is sumbitted (Trueman et al., submitted) and the uncertainties calculated here will be included in this complete analysis. 

For long $\gamma$ values, such as 31.6 Myr considered in Table~\ref{tab:ESSvaluesTiso}, the $^{107}$Pd/$^{108}$Pd and  $^{182}$Hf/$^{180}$Hf ratios would likely mostly reflect their production in one event only (regime III). In this case we derive an T$_\text{LE}$. 
Since $^{107}$Pd/$^{182}$Hf is between Regimes 1 and 3, this isotopic ratio changes more significantly during the time interval T$_\text{LE}$ than in the case of the \textit{r}-process isotopes discussed in the previous section. In the Table~\ref{tab:ESSvaluesTiso} we report the production value predicted by decaying back the ESS ratio by T$_\text{LE}$. As in the case of the \textit{r}-process isotopes, in this regimes this number can be used to determine the stellar yields of the last AGB star to have contributed to the \textit{s}-process elements present in the ESS (Trueman et al., submitted).

\subsection{The ratio of the \textit{p}-process $^{97}$Tc and $^{98}$Tc}

These two SLRs are next to each other in mass and are both \textit{p}-only isotopes, i.e., they are nuclei heavier than Fe that can only be produced by charged-particle reactions or the disintegration ($\gamma$) process. While the origin of \textit{p}-only isotopes is currently not well established especially for those in the light mass region, and the main sites may be both core-collapse and Type Ia supernovae, recent work has shown that the main site of production of the SLRs considered here is probably Chandrasekhar-mass Type Ia supernovae \citep[see, e.g.][]{Travaglio2014,Lugaro2016,Travaglio2018}. Because their mean lives are remarkably similar ($\tau$=5.94 and 6.1 Myr, respectively for $^{97}$Tc and $^{98}$Tc), their $\tau_\text{eq}$=226 Myr and as shown in Table \ref{tab:specificValuesSync}, the theoretical uncertainties related to their ratio are very low for values $\gamma$ up to 31.6 Myr. 

The full GCE of these isotopes was investigated by \citet{Travaglio2014}. Expanding on that work, in combination with the present results, could provide us with a strong opportunity to investigate both the origin of these \textit{p}-nuclei and the environment of the birth of the Sun. There are many scenarios that could in principle be investigated. If the $\gamma$ value of the origin Type Ia supernovae site was around 1 Myr, then we could derive a T$_\text{iso}$ from all the different ratios, and check for self-consistency. If the $\gamma$ value of the origin site was above 30 Myr, instead, we would be in a similar case as the \textit{r}-process isotopes discussed above, and the $^{97}$Tc/$^{98}$Tc would give us directly the production ratio in the original site, to be checked against nucleosynthesis predictions. For $\gamma$ values in-between, the $^{97}$Tc/$^{98}$Tc ratio would still provide us with the opportunity to calculate T$_\text{iso}$. Unfortunately, we only have upper limits for the ESS ratio of these two nuclei, relative to their experimental reference isotope $^{98}$Ru, which means that an ESS value for their ratio cannot be given and a detailed analysis needs to be postponed until such data becomes available.

\subsection{The ratio of $^{97}$Tc and $^{53}$Mn, also potentially of Chandrasekhar-mass Type Ia supernova origin}

From a chemical evolution perspective, the origin of Mn (and therefore $^{53}$Mn) is still unclear \citep{Seitenzahl13,Cescutti17,Eitner20,Kobayashi20,Lach20}.
Nevertheless, the $^{53}$Mn/$^{97}$Tc ratio can be assumed to be synchronous, as there are indications that the main site of origin of $^{53}$Mn is the same as that of $^{97}$Tc \footnote{And $^{98}$Tc, however, we prefer to consider $^{97}$Tc here because both its mean life and its yields are closer to that of $^{53}$Mn} \citep[see, e.g.,][]{Lugaro2016}.
Table~\ref{tab:specificValuesSync} shows that the uncertainty for the ratio of the two SLRs is below 30\% for most cases (and as low as 5\% when $\gamma$=1 Myr), while for each one of the individual isotopes is larger than 60\%. 
Similar to the $^{97}$Tc/$^{98}$Tc ratio discussed above, the $^{53}$Mn/$^{97}$Tc ratio can also provide the opportunity to investigate T$_\text{iso}$ for $\gamma$ values up to 2 Myr, because even if $\tau_\text{eq}$=59.4 Myr the shorter mean lives of each SLR do not allow to built a memory making this a case of Regime 3, which cannot be treated here. The ISM values reported in Table~\ref{tab:ESSvaluesTiso} were calculated with a production ratio of $2.39 \times 10^{-2}$ for $^{97}$Tc/$^{98}$Ru, $0.108$ for $^{53}$Mn/$^{55}$Mn, and $1.82 \times 10^{6}$ for $^{53}$Mn/$^{97}$Tc \citep{Lugaro2016,Travaglio2011}.

We obtain potential self-consistent isolation times, mostly determined by the accurate ESS value of $^{53}$Mn/$^{55}$Mn. Consistency between the last two columns of the table, which could inform us on the relative production of nuclei from nuclear statistical equilibrium (such as $^{53}$Mn) and nuclei from $\gamma$-process in Chandrasekhar-mass Type Ia supernovae (such as $^{97}$Tc), could be found only if the $^{97}$Tc/$^{98}$Ru ratio in the ESS was 7.3 times lower than the current upper limit. 

Similarly to the $s$-process case described above, for high values of $\gamma$ (e.g., 31.6 and 100 Myr shown in Table~\ref{tab:ESSvaluesTiso}), the $^{53}$Mn/$^{55}$Mn and $^{97}$Tc/$^{98}$Ru ratios would record one event only (Regime III) and the derived T$_\text{LE}$ are consistent with each other. The value from $^{53}$Mn/$^{55}$Mn can then be used to decay back the ESS ratio of the $^{53}$Mn/$^{97}$Tc and derive a direct constrain for the last \textit{p}-process event that polluted the solar material.
Overall, a more precise $^{97}$Tc ESS abundance would allow us to take advantage of the low theoretical uncertainties and give a more accurate prediction of the ISM ratio or the production ratio at the site.

\subsection{$^{60}$Fe/$^{26}$Al}

Finally we consider the case for $^{60}$Fe/$^{26}$Al. This ratio is of great interest in the literature because both isotopes are produced by core-collapse supernovae \citep{LimongiChieffi2006} and they can be observed with $\gamma$-rays \citep{Wang2007} as well as in the ESS \citep{Trappitsch2018}. There are strong discrepancies between core-collapse supernovae yields and observations, as the yields tipically produce a $^{60}$Fe/$^{26}$Al ratio at least three times higher than the $\gamma$-ray observations \citep[e.g.][]{Sukhbold2016}, and orders of magnitude higher than the ESS ratio \citep[see discusssion in][]{Lugaro2018}. 

We cannot apply our analysis to interpret the $\gamma$-ray ration because it is derived by measuring first the total abundance of $^{60}$Fe and $^{26}$Al separatedly, and then dividing them. In this case, the average abundance ratio is given simply by the ratio of the averages, mixing the $^{60}$Fe and $^{26}$Al productions from several different events, which do not correspond to our synchronous framework.

When considering the ESS abundance, however, we can apply our methods, since the ESS ratio represents abundance at one time and place in the ISM, generated by a synchronous set of events. In this case, $\tau_1 = 3.78$ Myr (for $^{60}$Fe) and $\tau_2 = 1.035$ Myr (for $^{26}$Al) results in $\tau_\text{eq} = -1.45$ Myr. If we consider a $\gamma = 1$ Myr for the core-collapse supernovae enriching events we fall somewhere between Regime 2 and 4, with $^{60}$Fe and $^{26}$Al building memory and almost no memory, respectively, between successive events. As a consequence, when considering our statistical analysis, the average ISM value given by Eq.~(\ref{eq:approxAvg}) predicted for the $^{60}$Fe/$^{26}$Al ratio is a factor of 3.9 of the production ratio. This is a $7\%$ higher than the traditional continuous enrichment steady-state formula $P \tau_1/\tau_2$ (i.e., the limit of Eq.~(\ref{eq:evolRatio}) when $\delta_c,\Delta t \to 0$) used in the literature \citep[see e.g.][]{Sukhbold2016}, since that gives a factor of 3.65 of the production ratio instead.
In conclusion, our analysis does not help to solve the problem that core-collapse supernova yields produce much more $^{60}$Fe relative to $^{26}$Al than observed in the ESS. 

\section{Conclusions and future work} \label{sec:conclusions}

We presented a statistical framework to study the uncertainties of ratios of SLRs that were present at the formation time of the Solar System. We show that this statistical framework is advantageous because: 
\begin{itemize}
\item it removes the GCE uncertainties associated with the stable reference isotopes often used for ESS ratios (i.e., the value of the parameter K investigated by \citealt{Cote2019}); 
\item it reduces the stochastic uncertainties, i.e., for ratios of two SLRs these uncertainties are typically much lower than those of SLR/stable isotopic ratios, for equivalent regimes.
\item it allows us to define a Regime 3 for the ratio of two SLRs, which is qualitatively different to the regimes described in Paper I for SLR/stable ratios, and represents the case where each mean life is much shorter than $\gamma$, while the equivalent mean life of the ratio of the two SLRs is much longer than $\gamma$. In this case the ratios of the two SLRs allows us to constrain the nucleosynthesis inside the last nucleosynthetic events that contributed the Solar System matter.
\end{itemize}

We have identified four ratios: $^{247}$Cm/$^{129}$I (from the \textit{r} process), $^{107}$Pd/$^{182}$Hf (from the \textit{s} process), $^{97}$Tc/$^{98}$Tc (from the \textit{p} process), and $^{53}$Mn/$^{97}$Tc (potentially from Type Ia supernovae), which can be used effectively to either reduce the uncertainty in the T$_\text{iso}$ calculation (for relatively small values of $\gamma$), or to predict accurately the production ratio for the last event that enriched the ESS (for relatively large values of $\gamma$). In particular, the inconsistencies we found (see Table~\ref{tab:ESSvaluesTiso}) between the production and the ESS ratios both for the $^{247}$Cm/$^{129}$I and the $^{107}$Pd/$^{182}$Hf ratios can be used to constrain the events in the Galaxy that produced the \textit{r}-process isotopes \citep{Cote2020} and the elements belonging to the first \textit{s}-process peak (Trueman et al., submitted) at the time of the formation of the Sun .

While here we have only investigated the simpler synchronous enrichment scenario, where the two SLRs are assumed to originate from the same events, in the future, we could also investigate the asynchronous enrichment scenario, for particular cases such as the $^{146}$Sm/$^{244}$Pu ratio. For example, $^{146}$Sm is a \textit{p} nucleus and $^{244}$Pu is produced by the \textit{r} process, therefore, the $\gamma$ for the production events of the two isotopes are probably very different. The mean life of $^{244}$Pu is 115 Myr, while for $^{146}$Sm, two different mean lives are reported: 98 Myr \citep{Kinoshita2012} and 149 Myr \citep{Marks2014}, for which $\tau_\text{eq} = 663$ Myr and $\tau_\text{eq} = 504$ Myr, respectively. Since these values are extremely long, the $^{146}$Sm/$^{244}$Pu ratio may provide with an opportunity to predict its value with an uncertainty much lower than when considering the individual isotopes. Another interesting ratio may be $^{135}$Cs/$^{60}$Fe, with a $\tau_\text{eq} = 26$ Myr (from mean lives of 3.3 and 3.78 Myr, respectively). For a frequent enrichment rate ($\gamma \sim 1$ Myr) the relative uncertainty on the predicted abundance ratio in a synchronous scenario is 4.5\%. However, $^{135}$Cs is a product of both the $s$ and the \textit{r} processes, while $^{60}$Fe is ejected mostly by core-collapse supernovae, which would require a complex asynchronous scenario. Furthermore, only an upper limit for the ESS abundance for $^{135}$Cs is available. 

In general, improvements in ESS data for any of the SLRs considered here will help us to constrain the stellar nucleosynthesis models. Particularly, these improvements are strongly needed for the \textit{p}-process isotopes $^{97}$Tc and $^{98}$Tc, for which we currently only have upper limits for their ESS abundances. Together with the well known $^{53}$Mn, these SLRs could provide unique constrains on both galactic \textit{p}-process nucleosynthesis and the origin of Solar System matter.

\acknowledgments
We thank the anonymous referee for the careful reading of the paper. This research is supported by the ERC Consolidator Grant (Hungary) funding scheme (Project RADIOSTAR, G.A. n. 724560). BC acknowledges the support from the National Science Foundation (NSF, USA) under grant No. PHY-1430152 (JINA Center for the Evolution of the Elements), and from the Hungarian Academy of Sciences via the Lend\"ulet project LP2014-17.

\vspace{5mm}

\bibliographystyle{aasjournal.bst}
\bibliography{aylrefs}

\appendix
\counterwithin{figure}{section}

\section{Calculation of $M_1/M_2$ average and standard deviation for the case when $\delta$ is not a constant}
\label{sec:mathDevelopment}

We can define the value of $M_1/M_2$ by parts as a function of time with
\begin{equation}
    \frac{M_1}{M_2} = P \sum_{i = 1}^n R_i(t),
\end{equation}
with
\begin{equation}
    R_i(t) = \frac{e^{-t/\tau_1} + \sum_{k = 0}^{i - 2}e^{-(t - \sum_{m = 0}^k\delta_m)/\tau_1}}{e^{-t/\tau_2} + \sum_{k = 0}^{i - 2}e^{-(t - \sum_{m = 0}^k\delta_m)/\tau_2}},
\end{equation}
where $\delta_{i - 1} < t$. Alternatively, the $R_i(t)$ can be written as
\begin{equation}
    R_i(t) = e^{-t/\tau_\text{eq}}C_i = e^{-t/\tau_\text{eq}}\frac{1 + \sum_{k = 0}^{i - 2}e^{\sum_{m = 0}^k\delta_m/\tau_1}}{1 + \sum_{k = 0}^{i - 2}e^{\sum_{m = 0}^k\delta_m/\tau_2}}.
    \label{eq:easyExpression}
\end{equation}

By using Equation~(\ref{eq:easyExpression}) with $P = 1$, the expression for the average value of $M_1/M_2$ is simply
\begin{equation}
    \left\langle\frac{M_1}{M_2}\right\rangle = \frac{1}{n\langle\delta\rangle}\sum_{i = 1}^n C_i \int_{\sum_{m = 0}^{i - 2}\delta_m}^{\sum_{m = 0}^{i - 1}\delta_m} e^{-t/\tau_\text{eq}} dt,
\end{equation}
from where
\begin{equation}
    \left\langle\frac{M_1}{M_2}\right\rangle = \frac{\tau_\text{eq}}{n\langle\delta\rangle}\sum_{i = 1}^n C_i \left(e^{-\sum_{m = 0}^{i - 2}\delta_m/\tau_\text{eq}} - e^{-\sum_{m = 0}^{i - 1}\delta_m/\tau_\text{eq}}\right).
\end{equation}

From the definition of $C_i$, we can rewrite the average as
\begin{equation}
    \left\langle\frac{M_1}{M_2}\right\rangle = \frac{\tau_\text{eq}}{n\langle\delta\rangle}\sum_{i = 1}^n \frac{1 + \sum_{k = 0}^{i - 2}e^{-\sum_{m = k}^{i - 2}\delta_m/\tau_1}}{1 + \sum_{k = 0}^{i - 2}e^{-\sum_{m = k}^{i - 2}\delta_m/\tau_2}}\left(1 - e^{-\delta_{i - 1}/\tau_\text{eq}}\right) = \frac{\tau_\text{eq}}{n\langle\delta\rangle}\sum_{i = 1}^n S_i \left(1 - e^{-\delta_{i - 1}/\tau_\text{eq}}\right).
    \label{eq:totAvg}
\end{equation}
Or, by taking the averages,
\begin{equation}
    \mu = \left\langle\frac{M_1}{M_2}\right\rangle = \frac{\tau_\text{eq}}{\langle\delta\rangle}\left(\langle S \rangle - \langle Se^{-\delta/\tau_\text{eq}} \rangle\right).
    \label{eq:correctAvg}
\end{equation}

We can obtain a more intuitive expression by approximating
\begin{equation}
    \langle S e^{-\delta/\tau_\text{eq}} \rangle \approx \langle S\rangle \langle e^{-\delta/\tau_\text{eq}} \rangle,
\end{equation}
and
\begin{equation}
    \langle S \rangle \approx \frac{1 - \langle e^{-\delta/\tau_2} \rangle}{1 - \langle e^{-\delta/\tau_1} \rangle},
\end{equation}
from where we can obtain that
\begin{equation}
    \mu = \left\langle\frac{M_1}{M_2}\right\rangle \approx \frac{\tau_\text{eq}}{\langle\delta\rangle}\frac{1 - \langle e^{-\delta/\tau_2}\rangle}{1 - \langle e^{-\delta/\tau_1}\rangle} \left(1 - \langle e^{-\delta/\tau_\text{eq}} \rangle \right).
\end{equation}

In order to calculate the standard deviation, we have to obtain the expression for the average of $(M_1/M_2)^2$. Following similar steps as before, we find that
\begin{equation}
    \left\langle\left(\frac{M_1}{M_2}\right)^2\right\rangle = \frac{\tau_\text{eq}}{2\langle\delta\rangle}\left(\langle S^2 \rangle - \langle S^2e^{-2\delta/\tau_\text{eq}} \rangle\right).
    \label{eq:correctAvgSq}
\end{equation}

From Equations (\ref{eq:correctAvg}) and (\ref{eq:correctAvgSq}), we can calculate the exact standard deviation as
\begin{equation}
    \sigma = \sqrt{\left\langle\left(\frac{M_1}{M_2}\right)^2\right\rangle - \left\langle\frac{M_1}{M_2}\right\rangle^2}.
    \label{eq:correctSigma}
\end{equation}

However, in the interest of having a more intuitive expression, we can approximate
\begin{equation}
    \langle S^2 \rangle \approx \frac{\left(1 + \langle e^{-\delta/\tau_1} \rangle\right)\left(1 - \langle e^{-2\delta/\tau_2}\rangle \right) \left(1 - \langle e^{-\delta/\tau_2} \rangle\right)}{\left(1 + \langle e^{-\delta/\tau_2}\rangle \right)\left(1 - \langle e^{-2\delta/\tau_1}\rangle\right)\left(1 - \langle e^{-\delta/\tau_1} \rangle \right)},
\end{equation}
from where we can get the final expression
\begin{equation}
    \frac{\sigma}{\mu} \approx \sqrt{\frac{\langle\delta\rangle}{2\tau_\text{eq}}\frac{1 - \langle e^{-2\delta/\tau_\text{eq}} \rangle}{\left(1 - \langle e^{-\delta/\tau_\text{eq}}\rangle\right)^2} - 1}.
    \label{eq:approxSigma1}
\end{equation}


\end{document}